\documentclass{svmult}

\usepackage{axodraw}
\usepackage{epsfig}

\def\bq{\begin{eqnarray}}
\def\eq{\end{eqnarray}}
\def\eps{\varepsilon}

\begin{document}


\title*{Algebraic Algorithms in Perturbative Calculations}
\author{Stefan Weinzierl}
\institute{Dipartimento di Fisica, Universit\`a di Parma,
INFN Gruppo Collegato di Parma, 43100 Parma, Italy,
\texttt{stefanw@fis.unipr.it}
}
%
%
\maketitle

I discuss algorithms for the evaluation of Feynman integrals. 
These algorithms are based on Hopf algebras and evaluate the
Feynman integral to (multiple) polylogarithms.

\section{Introduction}
\label{sect:intro}

Multiple polylogarithms are an object of interest not only for
mathematicians, but also for physicists in the domain of particle
physics.
Here, I will discuss how they occur in the calculation of Feynman
loop integrals.
The evaluation of these integrals is an essential part to 
obtain precise theoretical predictions on quantities, which can
be observed in experiment.
These predictions have a direct impact on searches for signals of
``new physics''.

Due to the complexity of these calculations, computer algebra plays an
essential part.
This in turn requires that methods developed to evaluate Feynman
loop integrals are suitable for an implementation on a computer.
The focus for the practitioner shifts therefore from the calculation
of a particular integral to the development of algorithms for a class
of integrals.
This shift is accompanied by a movement from concrete
analytical methods towards abstract algebraic algorithms.
In this article I review some algebraic techniques to solve
Feynman loop integrals.

In the next section I show briefly why loop corrections are needed
for today's experiments and how Feynman loop integrals 
arise in a pratical context.
Sect. \ref{sect:alg} introduces the algebraic tools:
A particular form of nested sums, which form a Hopf algebra
and which admits as additional structure a conjugation and
a convolution product.
In Sect. \ref{sect:hypergeom} I show how these tools are used
for the solution of a simple one-loop integral.
Special cases of the nested sums are multiple polylogarithms, which are 
discussed in Sect. \ref{sect:multipolylogs}. These multiple polylogarithms
admit a second Hopf algebra structure.
In Sect. \ref{sect:antipode} I discuss how the antipode of the Hopf algebra 
can be used to simplify expressions.

\section{Phenomenology}
\label{sect:pheno}

Phenomenology is the part of theoretical particle physics, which is 
most closely related to experiments.
The standard experiment in particle physics accelerates two particles, brings them into
collision and observes the outcome.
Examples for such experiments are Tevatron at Fermilab in Chicago, HERA at DESY in Hamburg
or LEP and the forthcoming LHC at CERN in Geneva.
Quite often a bunch of particles moving in one direction is observed.
This is called a (hadronic) jet and the direction of movement follows
closely the direction of the initial QCD partons (e.g. quarks and gluons),
which where created immediately after the collision.
The observed events can be classified according to their experimental 
signature, like the number of jets seen within one event.
Interesting questions related to these events are for example:
How often do we observe events with two or three jets ? 
What is the angular distribution of these multi-jet events~?
What is the value for specific observables like
the thrust, defined by 
\bq
T & = & \mbox{max}_{\vec{n}} 
        \frac{\sum\limits_i \left| \vec{p}_i \cdot \vec{n} \right| }
             {\sum\limits_i \left| \vec{p}_i \right| },
\eq
which maximizes the total longitudinal momentum 
of all final state particles $p_i$ along a unit vector $\vec{n}$ ?
There are many more interesting observables, and theoreticians 
in phenomenology try to provide predictions for those.
Obviously, it is desirable not to start a new calculation for each
observable, but to have a generic program, which provides predictions
for a wide class of observables.
This forces us to work with fully differential quantities.
This leaves in the main part of the calculation many kinematical
invariants, which cannot be integrated out.
As a general rule, the higher the number of jets is in an event, the more
complicated it is to obtain a theoretical prediction, since the number
of independent kinematical invariants increases.

By comparing the measurements with theoretical predictions
one can deduce information on the original
scattering process.
By counting the number of events with a particular signature 
one may discover new effects and
new particles.
However, it quite often occurs that hypothetical new particles lead to the
same signature in the detector as well known processes within the Standard
Model of particle physics.
To distinguish if a small measured excess in one observable is due to 
``new physics'' or only due to known physics requires precise theoretical prediction
from theory.
In the case where the expected events due to ``new physics'' are only a fraction
of the events from Standard Model physics, it is most important to have
a small theoretical uncertainty for the Standard Model background processes.
To give a simple example, if the number of events due to 
``new physics'' is about $1\%$ 
of the number of events for background processes within the Standard Model, 
then we need a theoretical prediction of the background with a precision
better than $1\%$.
At the energies where the experiments are done, all coupling constants are small
and perturbation theory in the coupling constants is the standard procedure
to obtain theoretical predictions.
Therefore to reduce the theoretical uncertainty on a prediction requires us 
to calculate higher orders in perturbation theory.
In phenomenology we are now moving towards fully differential
next-to-next-to-leading calculations, e.g. predictions which
include the first
three orders in the perturbative expansion.
\begin{figure}
\begin{center}
\begin{picture}(100,70)(0,30)
\Photon(30,50)(60,50){4}{4}
\Vertex(60,50){2}
\ArrowLine(60,50)(75,65)
\Line(75,65)(90,80)
\Line(90,20)(75,35)
\ArrowLine(75,35)(60,50)
\GlueArc(60,50)(20,-45,45){4}{4}
\Vertex(74,64){2}
\Vertex(74,36){2}
\Vertex(80,70){2}
\Gluon(80,70)(100,70){4}{2}
\Text(95,85)[l]{$p_1$}
\Text(105,70)[l]{$p_2$}
\Text(95,20)[l]{$p_3$}
\Vertex(30,50){2}
\ArrowLine(0,80)(30,50)
\ArrowLine(30,50)(0,20)
\Text(-5,20)[r]{$p_4$}
\Text(-5,85)[r]{$p_5$}
\end{picture} 
\end{center}
\caption{\label{fig1} A one-loop Feynman diagram contributing to the process
$e^+ e^- \rightarrow q g \bar{q}$.}
\end{figure}
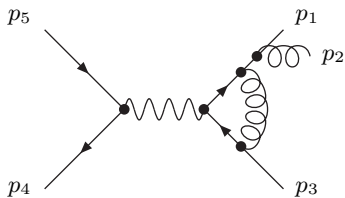    
An essential ingredient of higher order contributions are loop amplitudes.
Fig. \ref{fig1} shows a Feynman diagram contributing to the one-loop corrections
for the process $e^+ e^- \rightarrow q g \bar{q}$.
From the Feynman rules one obtains for this diagram (ignoring coupling and colour prefactors):
\bq
\label{feynmanrules}
- \bar{v}(p_4) \gamma^\mu v(p_5)
  \frac{1}{p_{123}^2}
  \int \frac{d^{4-2\eps}k_1}{(2\pi)^{4-2\eps}}
  \frac{1}{k_2^2}
  \bar{u}(p_1) \eps\!\!\!/(p_2) \frac{p\!\!\!/_{12}}{p_{12}^2}
  \gamma_\nu \frac{k\!\!\!/_1}{k_1^2}
  \gamma_\mu \frac{k\!\!\!/_3}{k_3^2}
  \gamma^\nu
  u(p_3).
\eq
Here, $p_{12}=p_1+p_2$, $p_{123}=p_1+p_2+p_3$, $k_2=k_1-p_{12}$, $k_3=k_2-p_3$.
Further $\eps\!\!\!/(p_2) = \gamma_\tau \eps^\tau(p_2)$, where $\eps^\tau(p_2)$ is the
polarization vector of the outgoing gluon.
All external momenta are assumed to be
massless: $p_i^2=0$ for $i=1..5$.
Dimensional regularization is used to regulate both ultraviolet and infrared
divergences.
In (\ref{feynmanrules}) the loop integral to be calculated reads
\bq
\label{exampleintegral}
  \int \frac{d^{4-2\eps}k_1}{(2\pi)^{4-2\eps}}
  \frac{k_1^\rho k_3^\sigma}{k_1^2 k_2^2 k_3^2}.
\eq
This loop integral contains the loop momentum $k_1$ in the numerator.
The further steps to evaluate this integral are now:
\begin{description}
\item[Step 1] Eliminate powers of the loop momentum in the numerator.
\item[Step 2] Convert the integral into an infinite sum.
\item[Step 3] Expand the sum into a Laurent series in $\eps$. 
\end{description}
The first two steps are rather easy to perform and convert the original
integral to a more convenient form.
The essential part is step 3.
It should be noted that the integral in (\ref{exampleintegral}) is rather simple and 
can be evaluated by other means.
However, I would like to discuss methods which generalize to higher loops and this
particular integral should be viewed as a pedagogical example.

To eliminate powers of the loop momentum in the numerator one can trade the loop
momentum in the numerator for scalar integrals (e.g. numerator $= 1$) with higher
powers of the propagators and shifted dimensions \cite{Tarasov:1996br,Tarasov:1997kx}:
\bq
\label{scalarintegral}
  \int \frac{d^{2m-2\eps}k_1}{(2\pi)^{2m-2\eps}}
  \frac{1}{\left(k_1^2\right)^{\nu_1} \left(k_2^2\right)^{\nu_2} \left(k_3^2\right)^{\nu_3}}.
\eq
This algorithm introduces temporarily Schwinger parameters together with raising
and lowering operators and expresses one integral of type (\ref{exampleintegral}) in terms of several
integrals of type (\ref{scalarintegral}).

In the second step an integral of type (\ref{scalarintegral}) is now converted into
an infinite sum. Introducing Feynman parameters, performing the momentum integration
and then the integration over the Feynman parameters one obtains
\bq
\label{integralresult}
\lefteqn{
 \int \frac{d^{2m-2\eps}k_1}{i \pi^{m-\eps}}
 \frac{1}{(-k_1^2)^{\nu_1}}
 \frac{1}{(-k_2^2)^{\nu_2}}
 \frac{1}{(-k_3^2)^{\nu_3}}
 }
 \nonumber \\
 & = & 
 \left( - p_{123}^2 \right)^{m-\eps-\nu_{123}}
 \frac{\Gamma(\nu_{123}-m+\eps)}{\Gamma(\nu_1)\Gamma(\nu_2)\Gamma(\nu_3)}
 \int\limits_0^1 da \; a^{\nu_2-1} (1-a)^{\nu_3-1}
 \nonumber \\
 & & 
 \times
 \int\limits_0^1 db \; b^{m-\eps-\nu_{23}-1} (1-b)^{m-\eps-\nu_1-1}
 \left[ 1- a \left(1-x\right) \right]^{m-\eps-\nu_{123}}
 \nonumber \\
 & = &
 \left( - p_{123}^2 \right)^{m-\eps-\nu_{123}}
 \frac{1}{\Gamma(\nu_1)\Gamma(\nu_2)}
 \frac{\Gamma(m-\eps-\nu_1)\Gamma(m-\eps-\nu_{23})}{\Gamma(2m-2\eps-\nu_{123})}
 \nonumber \\
 & & \times
 \sum\limits_{n=0}^\infty
 \frac{\Gamma(n+\nu_2)\Gamma(n-m+\eps+\nu_{123})}
      {\Gamma(n+1)\Gamma(n+\nu_{23})}
 \left(1-x\right)^n,
\eq
where $x=p_{12}^2/p_{123}^2$, $\nu_{23}=\nu_2+\nu_3$ and $\nu_{123}=\nu_1+\nu_2+\nu_3$.
To arrive at the last line of (\ref{integralresult})
one expands $\left[ 1- a \left(1-x\right) \right]^{m-\eps-\nu_{123}}$ according to
\bq
\left(1-z\right)^{-c} & = & 
 \frac{1}{\Gamma(c)} \sum\limits_{n=0}^\infty
 \frac{\Gamma(n+c)}{\Gamma(n+1)} z^n.
\eq
Then all Feynman parameter integrals are of the form
\bq
 \int\limits_0^1 da \; a^{\mu-1} (1-a)^{\nu-1} & = &
 \frac{\Gamma(\mu) \Gamma(\nu)}{\Gamma(\mu+\nu)}.
\eq
The infinite sum in the last line of (\ref{integralresult})
is a hypergeometric function, where the small parameter $\eps$ occurs in 
the Gamma-functions.

More complicated loop integrals yield additional classes of infinite sums.
The following types of infinite sums occur:
\begin{description}
\item[Type A:]
\bq
\label{type_A}
     \sum\limits_{i=0}^\infty 
       \frac{\Gamma(i+a_1)}{\Gamma(i+a_1')} ...
       \frac{\Gamma(i+a_k)}{\Gamma(i+a_k')}
       \; x^i
\eq
Up to prefactors the hypergeometric functions ${}_{J+1}F_J$ fall into this class.
The example discussed above is also contained in this class.
\item[Type B:]
\bq
\label{type_B}
\lefteqn{
\hspace*{-1cm}
     \sum\limits_{i=0}^\infty 
     \sum\limits_{j=0}^\infty 
       \frac{\Gamma(i+a_1)}{\Gamma(i+a_1')} ...
       \frac{\Gamma(i+a_k)}{\Gamma(i+a_k')}
       \frac{\Gamma(j+b_1)}{\Gamma(j+b_1')} ...
       \frac{\Gamma(j+b_l)}{\Gamma(j+b_l')}
 } \nonumber \\
 & &
 \times
       \frac{\Gamma(i+j+c_1)}{\Gamma(i+j+c_1')} ...
       \frac{\Gamma(i+j+c_m)}{\Gamma(i+j+c_m')}
       \; x^i y^j
\eq
An example for a function of this type is given by the first Appell function $F_1$.
\item[Type C:]
\bq
\label{type_C}
\lefteqn{
     \sum\limits_{i=0}^\infty 
     \sum\limits_{j=0}^\infty 
       \left( \begin{array}{c} i+j \\ j \\ \end{array} \right)
       \frac{\Gamma(i+a_1)}{\Gamma(i+a_1')} ...
       \frac{\Gamma(i+a_k)}{\Gamma(i+a_k')}
}
 \nonumber \\
 & &
 \times
       \frac{\Gamma(i+j+c_1)}{\Gamma(i+j+c_1')} ...
       \frac{\Gamma(i+j+c_m)}{\Gamma(i+j+c_m')}
       \; x^i y^j
\eq
Here, an example is given by the Kamp\'e de F\'eriet function $S_1$.
\item[Type D:]
\bq
\label{type_D}
\lefteqn{
\hspace*{-2cm}
     \sum\limits_{i=0}^\infty 
     \sum\limits_{j=0}^\infty 
       \left( \begin{array}{c} i+j \\ j \\ \end{array} \right)
       \frac{\Gamma(i+a_1)}{\Gamma(i+a_1')} ...
       \frac{\Gamma(i+a_k)}{\Gamma(i+a_k')}
       \frac{\Gamma(j+b_1)}{\Gamma(j+b_1')} ...
       \frac{\Gamma(j+b_l)}{\Gamma(j+b_l')}
 } \nonumber \\
 & &
 \times
       \frac{\Gamma(i+j+c_1)}{\Gamma(i+j+c_1')} ...
       \frac{\Gamma(i+j+c_m)}{\Gamma(i+j+c_m')}
       \; x^i y^j
\eq
An example for a function of this type is the second Appell function $F_2$.
\end{description}
Here, all $a_n$, $a_n'$, $b_n$, $b_n'$, $c_n$  and $c_n'$ are 
of the form ``integer $+ \;\mbox{const} \cdot \eps$''.
The task is now to expand these functions systematically into a Laurent series
in $\eps$, such that the resulting algorithms are suitable for an implementation
into a symbolic computer algebra program.
Due to the large size of intermediate expressions which occur in perturbative 
calculations, 
computer algebra systems like FORM \cite{Vermaseren:2000nd}
or GiNaC \cite{Bauer:2000cp}, 
which can handle large amounts of data
are used.
An implementation of the algorithms reviewed below can be found in \cite{Weinzierl:2002hv}.

\section{Nested Sums}
\label{sect:alg}

In this section I review the underlying mathematical structure
for the systematic expansion of the functions in (\ref{type_A})-(\ref{type_D}).
I discuss properties of particular forms of nested sums, 
which are called $Z$-sums and show that they form a Hopf algebra.
This Hopf algebra admits as additional structures 
a conjugation and a convolution product.
This summary is based on \cite{Moch:2001zr}, where additional information can
be found.
$Z$-sums are defined by
\bq 
\label{definition}
  Z(n;m_1,...,m_k;x_1,...,x_k) & = & \sum\limits_{n\ge i_1>i_2>\ldots>i_k>0}
     \frac{x_1^{i_1}}{{i_1}^{m_1}}\ldots \frac{x_k^{i_k}}{{i_k}^{m_k}}.
\eq
$k$ is called the depth of the $Z$-sum and $w=m_1+...+m_k$ is called the weight.
If the sums go to Infinity ($n=\infty$) the $Z$-sums are multiple polylogarithms \cite{Goncharov}:
\bq
\label{multipolylog}
Z(\infty;m_1,...,m_k;x_1,...,x_k) & = & \mbox{Li}_{m_k,...,m_1}(x_k,...,x_1).
\eq
For $x_1=...=x_k=1$ the definition reduces to the Euler-Zagier sums \cite{Euler,Zagier}:
\bq
Z(n;m_1,...,m_k;1,...,1) & = & Z_{m_1,...,m_k}(n).
\eq
For $n=\infty$ and $x_1=...=x_k=1$ the sum is a multiple $\zeta$-value \cite{Borwein}:
\bq
Z(\infty;m_1,...,m_k;1,...,1) & = & \zeta(m_k,...,m_1).
\eq
The multiple polylogarithms contain as the notation already suggests as subsets 
the classical polylogarithms 
$
\mbox{Li}_n(x)
$ 
\cite{lewin:book},
as well as
Nielsen's generalized polylogarithms \cite{Nielsen}
\bq
S_{n,p}(x) & = & \mbox{Li}_{1,...,1,n+1}(\underbrace{1,...,1}_{p-1},x),
\eq
and the harmonic polylogarithms \cite{Remiddi:1999ew}
\bq
\label{harmpolylog}
H_{m_1,...,m_k}(x) & = & \mbox{Li}_{m_k,...,m_1}(\underbrace{1,...,1}_{k-1},x).
\eq
The usefulness of the $Z$-sums lies in the fact, that they interpolate between
multiple polylogarithms and Euler-Zagier sums.

In addition to $Z$-sums, it is sometimes useful to introduce as well $S$-sums.
$S$-sums are defined by
\bq
S(n;m_1,...,m_k;x_1,...,x_k)  & = & 
\sum\limits_{n\ge i_1 \ge i_2\ge \ldots\ge i_k \ge 1}
\frac{x_1^{i_1}}{{i_1}^{m_1}}\ldots \frac{x_k^{i_k}}{{i_k}^{m_k}}.
\eq
The $S$-sums reduce for $x_1=...=x_k=1$ (and positive $m_i$) to harmonic sums \cite{Vermaseren:1998uu}:
\bq
S(n;m_1,...,m_k;1,...,1) & = & S_{m_1,...,m_k}(n).
\eq
The $S$-sums are closely related to the $Z$-sums, the difference being the upper summation boundary
for the nested sums: $(i-1)$ for $Z$-sums, $i$ for $S$-sums.
The introduction of $S$-sums is redundant, since $S$-sums can be expressed in terms
of $Z$-sums and vice versa.
It is however convenient to introduce both $Z$-sums and $S$-sums, since some 
properties are more naturally expressed in terms of
$Z$-sums while others are more naturally expressed in terms of $S$-sums.
An algorithm for the conversion from $Z$-sums to $S$-sums and vice versa can
be found in \cite{Moch:2001zr}.

The $Z$-sums form an algebra.
The unit element in the algebra is given by the empty sum
\bq
e & = & Z(n).
\eq
The empty sum $Z(n)$ equals $1$ for non-negative integer $n$.
Before I discuss the multiplication rule, let me note that the basic
building blocks of $Z$-sums are expressions of the form
\bq
\frac{x_j^n}{n^{m_j}},
\eq
which will be called ``letters''.
For fixed $n$, one can multiply two letters with the same $n$:
\bq
\label{alphabet}
\frac{x_1^n}{n^{m_1}} \cdot \frac{x_2^n}{n^{m_2}}
 & = & 
\frac{\left(x_1 x_2\right)^n}{n^{m_1+m_2}},
\eq
e.g. the $x_j$'s are multiplied and the degrees are added.
Let us call the set of all letters the alphabet $A$. 
As a short-hand notation I will in the following denote a letter just 
by $X_j=x_j^n/n^{m_j}$. 
A word is an ordered sequence of letters, e.g.
\bq
W & = & X_1, X_2, ..., X_k.
\eq
The word of length zero is denoted by $e$.
The $Z$-sums defined in (\ref{definition}) are therefore completely
specified by the upper summation limit $n$ 
and a word $W$.
A quasi-shuffle algebra ${\cal A}$ on the vectorspace of words 
is defined by \cite{Hoffman} 
\bq 
\label{algebra}
e \circ W & = & W \circ e = W, \nonumber \\
(X_1,W_1) \circ (X_2,W_2) & = & X_1,(W_1 \circ (X_2,W_2)) + X_2,((X_1,W_1) \circ W_2) \nonumber \\
& & + (X_1 \cdot X_2),(W_1 \circ W_2).
\eq
Note that ``$\cdot$'' denotes multiplication of letters as defined 
in eq. (\ref{alphabet}),
whereas ``$\circ$'' denotes the product in the algebra ${\cal A}$, recursively
defined in eq. (\ref{algebra}).
This defines a quasi-shuffle product for $Z$-sums.
The recursive definition in (\ref{algebra}) translates for $Z$-sums into
\bq
\label{Zmultiplication}
\lefteqn{
Z_{m_1,...,m_k}(n) \times Z_{m_1',...,m_l'}(n) } & & \nonumber \\
& = & \sum\limits_{i_1=1}^n \frac{1}{i_1^{m_1}} Z_{m_2,...,m_k}(i_1-1) Z_{m_1',...,m_l'}(i_1-1) \nonumber \\
&  & + \sum\limits_{i_2=1}^n \frac{1}{i_2^{m_1'}} Z_{m_1,...,m_k}(i_2-1) Z_{m_2',...,m_l'}(i_2-1) \nonumber \\
&  & + \sum\limits_{i=1}^n \frac{1}{i^{m_1+m_1'}} Z_{m_2,...,m_k}(i-1) Z_{m_2',...,m_l'}(i-1).
\eq
\begin{figure}
\begin{center}
\epsfig{file=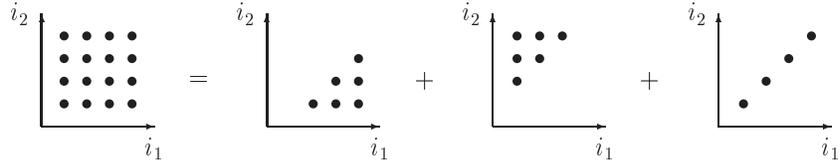, bbllx=100pt, bblly=600pt, bburx=500pt, bbury=700pt, width=12cm}
\caption{\label{proof} Sketch of the proof for the multiplication of $Z$-sums. The sum over the square is replaced by
the sum over the three regions on the r.h.s.}
\end{center}
\end{figure}
The proof that $Z$-sums obey the quasi-shuffle algebra is sketched in Fig. \ref{proof}.
The outermost sums of the $Z$-sums on the l.h.s of (\ref{Zmultiplication}) are split into the three
regions indicated in Fig. \ref{proof}.
A simple example for the multiplication of two $Z$-sums is 
\bq
\lefteqn{
Z(n;m_1;x_1) Z(n;m_1;x_2)  = }
\nonumber \\
 & & 
 Z(n;m_1,m_2;x_1,x_2) 
+Z(n;m_2,m_1;x_2,x_1)
+Z(n;m_1+m_2;x_1 x_2).
\eq

The quasi-shuffle algebra ${\cal A}$ is isomorphic to the free polynomial algebra on the Lyndon words.
If one introduces a lexicographic ordering on the letters of the alphabet
$A$, a Lyndon word is defined by the property
\bq
W < V
\eq
for any subwords $U$ and $V$ such that $W=U, V$.
Here $U, V$ means just concatenation of $U$ and $V$. 

The $Z$-sums form actuall a Hopf algebra.
It is convenient to phrase the coalgebra structure in terms of rooted trees.
$Z$-sums can be represented as rooted trees without any sidebranchings. 
As a concrete example the pictorial representation of a sum 
of depth three reads:
\\
\vspace*{-15mm}
\bq
Z(n;m_1,m_2,m_3;x_1,x_2,x_3) 
& = &
\sum\limits_{i_1=1}^n
\sum\limits_{i_2=1}^{i_1-1}
\sum\limits_{i_3=1}^{i_2-1}
 \frac{x_1^{i_1}}{{i_1}^{m_1}}
 \frac{x_2^{i_2}}{{i_2}^{m_2}}
 \frac{x_3^{i_3}}{{i_3}^{m_3}}
\;\; =  \;\;
\begin{picture}(40,60)(-10,30)
\Vertex(10,50){2}
\Vertex(10,30){2}
\Vertex(10,10){2}
\Line(10,10)(10,50)
\Text(6,50)[r]{$x_1$}
\Text(6,30)[r]{$x_2$}
\Text(6,10)[r]{$x_3$}
\end{picture} 
\eq
\\
\\
The outermost sum corresponds to the root. By convention, the root is always drawn
on the top.
Trees with sidebranchings are given by nested sums with more than one subsum, for example:
\\
\vspace*{-15mm}
\bq
\sum\limits_{i=1}^n \frac{x_1^i}{i^{m_1}} Z(i-1;m_2,x_2) Z(i-1;m_3;x_3) 
& = &
\vspace*{-8mm}
\begin{picture}(60,60)(0,30)
\Vertex(30,50){2}
\Vertex(10,20){2}
\Vertex(50,20){2}
\Line(30,50)(10,20)
\Line(30,50)(50,20)
\Text(26,50)[r]{$x_1$}
\Text(6,20)[r]{$x_2$}
\Text(46,20)[r]{$x_3$}
\end{picture}
\eq
\\
Of course, due to the multiplication formula, trees with sidebranchings can always be
reduced to trees without any sidebranchings.
The coalgebra structure is now formulated in terms of rooted trees.
I first introduce some notation how to manipulate rooted trees,
following the notation of Kreimer and Connes \cite{Kreimer:1998dp, Connes:1998qv}.
An elementary cut of a rooted tree is a cut at a single chosen edge. 
An admissible cut is any assignment of elementary
cuts to a rooted tree such that any path from any vertex of the tree to the root has at most one elementary cut.
An admissible cut maps a tree $t$ to a monomial in trees 
$t_1 \circ ... \circ t_{k+1}$. 
Note that precisely one of 
these subtrees $t_j$
will contain the root of $t$. Denote this distinguished tree by $R^C(t)$, and the monomial delivered by the $k$ other factors
by $P^C(t)$. The counit $\bar{e}$ is given by
\bq
\bar{e}(e) & = & 1, \nonumber \\
\bar{e}(t) & = & 0, \;\;\; t \neq e.
\eq
The coproduct $\Delta$ is defined by the equations
\bq
\label{defcoproduct}
\Delta(e) & = & e \otimes e, \nonumber \\
\Delta(t) & = & e \otimes t + t \otimes e + \sum\limits_{\mbox{\tiny adm. cuts $C$ of $t$}} P^C(t) \otimes R^C(t), \nonumber \\
\Delta(t_1 \circ ... \circ t_k ) & = & \Delta(t_1) ( \circ \otimes \circ ) ... ( \circ \otimes \circ ) \Delta(t_k).
\eq
The antipode ${\cal S}$ is given by
\bq
\label{defantipode}
{\cal S}(e) & = & e, \nonumber \\
{\cal S}(t) & = & -t - \sum\limits_{\mbox{\tiny adm. cuts $C$ of $t$}} {\cal S}\left( P^C(t) \right) \circ R^C(t), \nonumber \\
{\cal S}(t_1 \circ ... \circ t_k) & = & {\cal S}(t_1) \circ ... \circ {\cal S}(t_k).
\eq
Since the multiplication in the algebra is commutative the antipode satisfies
\bq
 {\cal S}^2 & = & \mbox{id}.
\eq
Let me give some examples for the coproduct and the antipode for $Z$-sums:
\bq
\Delta Z(n;m_1;x_1) & = & 
 e \otimes Z(n;m_1;x_1) + Z(n;m_1;x_1) \otimes e, \nonumber \\
\Delta Z(n;m_1,m_2;x_1,x_2) & = &
 e \otimes Z(n;m_1,m_2;x_1,x_2) + Z(n;m_1,m_2;x_1,x_2) \otimes e \nonumber \\ 
 & & 
 + Z(n;m_2;x_2) \otimes Z(n;m_1;x_1),
\eq
\bq
{\cal S} Z(n;m_1;x_1) & = & 
 - Z(n;m_1;x_1), \nonumber \\
{\cal S} Z(n;m_1,m_2;x_1,x_2) & = &
 Z(n;m_2,m_1;x_2,x_1) + Z(n;m_1+m_2;x_1 x_2).
\eq
The Hopf algebra of nested sums has additional structures if we allow expressions
of the form
\bq
\label{augmented}
\frac{x_0^n}{n^{m_0}} Z(n;m_1,...,m_k;x_1,...,x_k),
\eq
e.g. $Z$-sums multiplied by a letter.
Then the following convolution product
\bq
\label{convolution}
 \sum\limits_{i=1}^{n-1} \; \frac{x^i}{i^m} Z(i-1;...)
                         \; \frac{y^{n-i}}{(n-i)^{m'}} Z(n-i-1;...)
\eq
can again be expressed in terms of expressions of the form (\ref{augmented}).
An example is
\bq
\lefteqn{
 \sum\limits_{i=1}^{n-1} \; \frac{x^i}{i} Z_1(i-1)
                         \; \frac{y^{n-i}}{(n-i)} Z_1(n-i-1)
 = }
 \nonumber \\
 & &
 \frac{x^n}{n} \left[ 
    Z\left(n-1;1,1,1;\frac{y}{x},\frac{x}{y},\frac{y}{x}\right)
   +Z\left(n-1;1,1,1;\frac{y}{x},1,\frac{x}{y}\right)
 \right. \nonumber \\
 & & \left.
   +Z\left(n-1;1,1,1;1,\frac{y}{x},1\right)
 \right]
 + \left( x \leftrightarrow y \right).
\eq
In addition there is a conjugation, e.g. sums of the form 
\bq
\label{conjugation}
 - \sum\limits_{i=1}^n 
       \left( \begin{array}{c} n \\ i \\ \end{array} \right)
       \left( -1 \right)^i
       \; \frac{x^i}{i^m} S(i;...)
\eq
can also be reduced to terms of the form (\ref{augmented}).
Although one can easily convert between the notations for $S$-sums and
$Z$-sums, expressions involving a conjugation tend to be shorter when
expressed in terms of $S$-sums.
The name conjugation stems from the following fact:
To any function $f(n)$ of an integer variable $n$ one can define
a conjugated function $C \circ f(n)$ as the following sum
\bq
C \circ f(n) & = & \sum\limits_{i=1}^n 
       \left( \begin{array}{c} n \\ i \\ \end{array} \right)
       (-1)^i f(i).
\eq
Then conjugation satisfies the following two properties:
\bq
C \circ 1 & = & 1,
 \nonumber \\
C \circ C \circ f(n) & = & f(n).
\eq
An example for a sum involving a conjugation is
\bq
\lefteqn{
\hspace*{-1cm}
 - \sum\limits_{i=1}^n 
       \left( \begin{array}{c} n \\ i \\ \end{array} \right)
       \left( -1 \right)^i
       \; \frac{x^i}{i} S_1(i)
 = } \nonumber \\
 & &
 S\left(n;1,1;1-x, \frac{1}{1-x}\right)
 -S\left(n;1,1;1-x, 1\right).
\eq
Finally there is the combination of conjugation and convolution,
e.g. sums of the form 
\bq
\label{conjugationconvolution}
 - \sum\limits_{i=1}^{n-1} 
       \left( \begin{array}{c} n \\ i \\ \end{array} \right)
       \left( -1 \right)^i
       \; \frac{x^i}{i^m} S(i;...)
       \; \frac{y^{n-i}}{(n-i)^{m'}} S(n-i;...)
\eq
can also be reduced to terms of the form (\ref{augmented}).
An example is given by
\bq
\lefteqn{
 - \sum\limits_{i=1}^{n-1} 
       \left( \begin{array}{c} n \\ i \\ \end{array} \right)
       \left( -1 \right)^i
       \; S(i;1;x)
       \; S(n-i;1;y) =
} \nonumber \\
 & &
 \frac{1}{n} 
 \left\{
 S(n;1;y)
 + (1-x)^n 
   \left[ S\left(n;1;\frac{1}{1-\frac{1}{x}}\right)
        - S\left(n;1;\frac{1-\frac{y}{x}}{1-\frac{1}{x}}\right)
   \right]
 \right\}
 \nonumber \\
 & &
 + \frac{(-1)^n}{n}
 \left\{
 S(n;1;x)
 + (1-y)^n 
   \left[ S\left(n;1;\frac{1}{1-\frac{1}{y}}\right)
        - S\left(n;1;\frac{1-\frac{x}{y}}{1-\frac{1}{y}}\right)
   \right]
 \right\}.
 \nonumber \\
\eq

\section{Expansion of hypergeometric functions}
\label{sect:hypergeom}

In this section I discuss how the algebraic tools introduced in the previous
section can be used to solve the problems outlined at the end of Sect. \ref{sect:pheno}.
First I give some motivation for the introduction of $Z$-sums: The essential point is that
$Z$-sums interpolate between multiple polylogarithms and Euler-Zagier-sums, such that
the interpolation is compatible with the algebra structure.
On the one hand, we expect multiple polylogarithm to appear in the Laurent expansion
of the transcendental functions (\ref{type_A})-(\ref{type_D}), a fact which is confirmed
a posteriori. Therefore it is important that multiple polylogarithms are contained in the class
of $Z$-sums.
On the other the expansion parameter $\eps$ occurs in the functions
(\ref{type_A})-(\ref{type_D}) inside the arguments of Gamma-functions.
The basic formula for the expansion of Gamma-functions reads
\bq
\label{expansiongamma}
\lefteqn{
\hspace*{-1cm}
\Gamma(n+\eps)  = \Gamma(1+\eps) \Gamma(n)
 \left[
        1 + \eps Z_1(n-1) + \eps^2 Z_{11}(n-1)
 \right.
} \nonumber \\
 & & \left.
          + \eps^3 Z_{111}(n-1) + ... + \eps^{n-1} Z_{11...1}(n-1)
 \right],
\eq
containing Euler-Zagier sums for finite $n$.
As a simple example I discuss the expansion of 
\bq
\label{examplehypergeom}
\sum\limits_{i=0}^\infty
 \frac{\Gamma(i+a_1+t_1\eps)\Gamma(i+a_2+t_2\eps)}{\Gamma(i+1)\Gamma(i+a_3+t_3\eps)}
 x^i
\eq
into a Laurent series in $\eps$. Here $a_1$, $a_2$ and $a_3$ are assumed to be integers.
Up to prefactors the expression in (\ref{examplehypergeom}) is a hypergeometric function ${}_2F_1$.
Using $\Gamma(x+1) = x \Gamma(x)$, partial fractioning and an adjustment of the
summation index one can transform (\ref{examplehypergeom}) into terms of the form
\bq
\sum\limits_{i=1}^\infty
 \frac{\Gamma(i+t_1\eps)\Gamma(i+t_2\eps)}{\Gamma(i)\Gamma(i+t_3\eps)}
 \frac{x^i}{i^m},
\eq
where $m$ is an integer.
Now using (\ref{expansiongamma})
one obtains
\bq
\Gamma(1+\eps) 
\sum\limits_{i=1}^\infty
 \frac{\left(1+\eps t_1 Z_1(i-1)+...\right) \left(1+\eps t_2 Z_1(i-1)+...\right)}
      {\left(1+\eps t_3 Z_1(i-1)+...\right)}
 \frac{x^i}{i^m}.
\eq
Inverting the power series in the denominator and truncating in $\eps$ one obtains
in each order in $\eps$ terms of the form
\bq
\label{exZ1}
\sum\limits_{i=1}^\infty
 \frac{x^i}{i^m}
 Z_{m_1 ... m_k}(i-1) Z_{m_1' ... m_l'}(i-1) Z_{m_1'' ... m_n''}(i-1)
\eq
Using the quasi-shuffle product for $Z$-sums the three Euler-Zagier sums
can be reduced to single Euler-Zagier sums and one finally arrives at terms of the form
\bq
\label{exZ2}
\sum\limits_{i=1}^\infty
 \frac{x^i}{i^m}
 Z_{m_1 ... m_k}(i-1),
\eq
which are harmonic polylogarithms $H_{m,m_1,...,m_k}(x)$.
This completes the algorithm for the expansion in $\eps$ for sums of the form (\ref{type_A}).
Since the one-loop integral discussed in (\ref{integralresult}) 
is a special case of (\ref{type_A}),
this algorithm also applies to the integral (\ref{integralresult}).
In addition, this algorithm shows that in the expansion of hypergeometric functions
${}_{J+1}F_J(a_1,...,a_{J+1};b_1,...,b_J;x)$ around integer values of the parameters
$a_k$ and $b_l$  only harmonic polylogarithms appear in the result.

The algorithm for the expansion of sums of type (\ref{type_A}) used
the multiplication formula for $Z$-sums to pass from (\ref{exZ1}) to (\ref{exZ2}).
To expand double sums of type (\ref{type_B}) one needs in addition the 
convolution product (\ref{convolution}). 
To expand sums of type (\ref{type_C}) the conjugation (\ref{conjugation}) is needed.
Finally, for sums of type (\ref{type_D}) the combination of conjugation and convolution
as in (\ref{conjugationconvolution}) is required.
More details can be found in \cite{Moch:2001zr}.

Let me come back to the example of the one-loop Feynman integral discussed in 
Sect. \ref{sect:pheno}.
For $\nu_1=\nu_2=\nu_3=1$ and $m=2$ in (\ref{integralresult}) one obtains:
\bq
\lefteqn{
 \int \frac{d^{4-2\eps}k_1}{i \pi^{2-\eps}}
 \frac{1}{(-k_1^2)}
 \frac{1}{(-k_2^2)}
 \frac{1}{(-k_3^2)}
 }
 \nonumber \\
 & = & 
  \frac{\Gamma(-\eps)\Gamma(1-\eps)\Gamma(1+\eps)}{\Gamma(1-2\eps)}
 \frac{\left( - p_{123}^2 \right)^{-1-\eps}}{1-x}
 \sum\limits_{n=1}^\infty
 \eps^{n-1}
 H_{\underbrace{1,...,1}_{n}}(1-x).
\eq
Here, all harmonic polylogarithms can be expressed in terms of Nielsen
polylogarithms, which in turn simplify to powers of the standard logarithm:
\bq
H_{\underbrace{1,...,1}_{n}}(1-x) & = & 
 S_{0,n}(1-x)
 = \frac{(-1)^n}{n!} \left( \ln x \right)^n.
\eq
This particular example is very simple and one recovers the well-known
all-order result
\bq
  \frac{\Gamma(1-\eps)^2\Gamma(1+\eps)}{\Gamma(1-2\eps)}
 \frac{\left( - p_{123}^2 \right)^{-1-\eps}}{\eps^2}
 \frac{1-x^{-\eps}}{1-x},
\eq
which (for this simple example)
can also be obtained by direct integration. 
 
\section{Multiple polylogarithms}
\label{sect:multipolylogs}

The multiple polylogarithms are special cases of $Z$-sums.
They are obtained from $Z$-sums by taking the outermost sum to infinity:
\bq
Z(\infty;m_1,...,m_k;x_1,...,x_k) & = & \mbox{Li}_{m_k,...,m_1}(x_k,...,x_1).
\eq
The reversed order of the arguments and indices on the r.h.s. follows the notation
of Goncharov \cite{Goncharov}.
They have been studied extensively in the literature
by physicists
\cite{Remiddi:1999ew,Vermaseren:1998uu}, \cite{Gehrmann:2000zt}-\cite{Moch:2002hm}
and mathematicians
\cite{Borwein},\cite{Hain}-\cite{Racinet:2002}.
Here I summarize the most important properties.
Being special cases of $Z$-sums they obey the quasi-shuffle Hopf algebra for
$Z$-sums.
Multiple polylogarithms have been defined in this article via the sum representation
(\ref{multipolylog}).
In addition, they admit an integral representation. From this integral representation
a second algebra structure arises, which turns out to be a shuffle Hopf algebra.
To discuss this second Hopf algebra it is convenient to 
introduce for $z_k \neq 0$
the following functions
\bq
\label{Gfuncdef}
G(z_1,...,z_k;y) & = &
 \int\limits_0^y \frac{dt_1}{t_1-z_1}
 \int\limits_0^{t_1} \frac{dt_2}{t_2-z_2} ...
 \int\limits_0^{t_{k-1}} \frac{dt_k}{t_k-z_k}.
\eq
In this definition 
one variable is redundant due to the following scaling relation:
\bq
G(z_1,...,z_k;y) & = & G(x z_1, ..., x z_k; x y)
\eq
If one further defines
\bq
g(z;y) & = & \frac{1}{y-z},
\eq
then one has
\bq
\frac{d}{dy} G(z_1,...,z_k;y) & = & g(z_1;y) G(z_2,...,z_k;y)
\eq
and
\bq
\label{Grecursive}
G(z_1,z_2,...,z_k;y) & = & \int\limits_0^y dt \; g(z_1;t) G(z_2,...,z_k;t).
\eq
One can sligthly enlarge the set and define
$G(0,...,0;y)$ with $k$ zeros for $z_1$ to $z_k$ to be
\bq
\label{trailingzeros}
G(0,...,0;y) & = & \frac{1}{k!} \left( \ln y \right)^k.
\eq
This permits us to allow trailing zeros in the sequence
$(z_1,...,z_k)$ by defining the function $G$ with trailing zeros via (\ref{Grecursive}) 
and (\ref{trailingzeros}).
To relate the multiple polylogarithms to the functions $G$ it is convenient to introduce
the following short-hand notation:
\bq
\label{Gshorthand}
G_{m_1,...,m_k}(z_1,...,z_k;y)
 & = &
 G(\underbrace{0,...,0}_{m_1-1},z_1,...,z_{k-1},\underbrace{0...,0}_{m_k-1},z_k;y)
\eq
Here, all $z_j$ for $j=1,...,k$ are assumed to be non-zero.
One then finds
\bq
\label{Gintrepdef}
\mbox{Li}_{m_k,...,m_1}(x_k,...,x_1)
& = & (-1)^k 
 G_{m_1,...,m_k}\left( \frac{1}{x_1}, \frac{1}{x_1 x_2}, ..., \frac{1}{x_1...x_k};1 \right).
\eq
The inverse formula reads
\bq
G_{m_1,...,m_k}(z_1,...,z_k;y) & = & 
 (-1)^k \; \mbox{Li}_{m_k,...,m_1}\left(\frac{z_{k-1}}{z_k},...,\frac{z_1}{z_2},\frac{y}{z_1}\right).
\eq
Eq. (\ref{Gintrepdef}) together with 
(\ref{Gshorthand}) and (\ref{Gfuncdef})
defines an integral representation for the multiple polylogarithms.
To make this more explicit I first introduce some notation for iterated integrals
\bq
\int\limits_0^\Lambda \frac{dt}{t-a_n} \circ ... \circ \frac{dt}{t-a_1} & = & 
\int\limits_0^\Lambda \frac{dt_n}{t_n-a_n} \int\limits_0^{t_n} \frac{dt_{n-1}}{t_{n-1}-a_{n-1}} \times ... \times \int\limits_0^{t_2} \frac{dt_1}{t_1-a_1}
\eq
and the short hand notation:
\bq
\int\limits_0^\Lambda \left( \frac{dt}{t} \circ \right)^{m} \frac{dt}{t-a}
& = & 
\int\limits_0^\Lambda 
\underbrace{\frac{dt}{t} \circ ... \frac{dt}{t}}_{m \;\mbox{times}} \circ \frac{dt}{t-a}.
\eq
The integral representation for $\mbox{Li}_{m_k,...,m_1}(x_k,...,x_1)$ reads then
\bq
\label{intrepII}
\lefteqn{
\mbox{Li}_{m_k,...,m_1}(x_k,...,x_1) = 
 (-1)^k \int\limits_0^1 \left( \frac{dt}{t} \circ \right)^{m_1-1} \frac{dt}{t-b_1} 
 } \nonumber \\
 & & 
 \circ \left( \frac{dt}{t} \circ \right)^{m_2-1} \frac{dt}{t-b_2}
 \circ ... \circ
 \left( \frac{dt}{t} \circ \right)^{m_k-1} \frac{dt}{t-b_k},
\eq
where the $b_j$'s are related to the $x_j$'s 
\bq
b_j & = & \frac{1}{x_1 x_2 ... x_j}.
\eq
From the iterated integral representation (\ref{Gfuncdef}) 
a second algebra structure for the functions
$G(z_1,...,z_k;y)$ (and through (\ref{Gintrepdef}) also for the multiple polylogarithms)
is obtained as follows:
We take the $z_j$'s as letters and call a sequence of ordered letters $w=z_1,...,z_k$ a word.
Then the function $G(z_1,...,z_k;y)$ is uniquely specified by the word
$w=z_1,...,z_k$ and the variable $y$.
The neutral element $e$ is given by the empty word, equivalent to
\bq
G(;y) & = & 1.
\eq
A shuffle algebra on the vector space of words is defined by
\bq 
\label{defshuffleproduct}
e \circ w & = & w \circ e = w, \nonumber \\
(z_1,w_1) \circ (z_2,w_2) & = & z_1,(w_1 \circ (z_2,w_2)) + z_2,((z_1,w_1) \circ w_2).
\eq
Note that this definition is very similar to the definition of the quasi-shuffle algebra
(\ref{algebra}), except that the third term in (\ref{algebra}) is missing.
In fact, a shuffle algebra is a special case of a quasi-shuffle algebra, where
the product of two letters is degenerate: $X_1 \cdot X_2 = 0$ for all letters $X_1$ and $X_2$
in the notation of Sect. \ref{sect:alg}.
The definition of the shuffle product (\ref{defshuffleproduct}) translates into 
the following recursive definition of the product of two $G$-functions:
\bq
\lefteqn{
G(z_1,...,z_k;y) \times G(z_{k+1},...,z_n;y) = } \\
 & & 
 \int\limits_0^y \frac{dt}{t-z_1} G(z_2,...,z_k;t) G(z_{k+1},...,z_n;t) 
 \nonumber \\
 & &
 + \int\limits_0^y \frac{dt}{t-z_{k+1}} G(z_1,...,z_k;t) G(z_{k+2},...,z_n;t)
\eq

For the discussion of the coalgebra part for the functions $G(z_1,...,z_k;y)$ we may 
proceed as in Sect. \ref{sect:alg} and associate to any function $G(z_1,...,z_k;y)$ a rooted
tree without sidebranchings as in the following example:
\\
\vspace*{-15mm}
\bq
G(z_1,z_2,z_3;y)
& = &
\begin{picture}(40,60)(-10,30)
\Vertex(10,50){2}
\Vertex(10,30){2}
\Vertex(10,10){2}
\Line(10,10)(10,50)
\Text(6,50)[r]{$z_1$}
\Text(6,30)[r]{$z_2$}
\Text(6,10)[r]{$z_3$}
\end{picture} 
\eq
\\
\\
The outermost integration (involving $z_1$) corresponds to the root.
The formulae for the coproduct (\ref{defcoproduct}) and the 
antipode (\ref{defantipode}) apply then also to the functions
$G(z_1,...,z_k;y)$.

A shuffle algebra is simpler than a quasi-shuffle algebra and one finds for
a shuffle algebra besides the recursive definitions of the product, the coproduct
and the antipode also closed formulae for these operations.
For the product one has
\bq
G(z_1,...,z_k;y) \; G(z_{k+1},...,z_{k+l};y) 
& = & \sum\limits_{shuffle} G\left( z_{\sigma(1)}, ..., z_{\sigma(k+l)}; y \right),
\eq
where the sum is over all permutations which preserve the relative order of the strings
$z_1,...,z_k$ and $z_{k+1},...,z_{k+l}$.
This explains the name ``shuffle product''.
For the coproduct one has
\bq
\Delta G(z_1,...,z_k;y) & = & 
 \sum\limits_{j=0}^k G(z_1,...,z_j;y) \otimes G(z_{j+1},...,z_k;y)
\eq
and for the antipode one finds
\bq
\label{Gantipodeexpli}
{\cal S} G(z_1,...,z_k;y) & = & (-1)^k G(z_k,...,z_1;y).
\eq
The shuffle multiplication is commutative and the antipode satisfies therefore
\bq
 {\cal S}^2 & = & \mbox{id}.
\eq
From (\ref{Gantipodeexpli}) this is evident.

\section{The antipode and integration-by-parts}
\label{sect:antipode}

Integration-by-parts has always been a powerful tool for calculations
in particle physics.
By using integration-by-parts one may obtain an identity between various
$G$-functions. The starting point is as follows:
\bq
\lefteqn{
G(z_1,...,z_k;y) 
 =
 \int\limits_0^y dt \left( \frac{\partial}{\partial t} G(z_1;t) \right)
   G(z_2,...,z_k;y) }
 \nonumber \\
 & &
 =
  G(z_1;y) G(z_2,...,z_k;y) - \int\limits_0^y dt \; G(z_1;t) 
       g(z_2;t) G(z_3,...,z_k;y)
 \nonumber \\ 
 & &
 =
  G(z_1;y) G(z_2,...,z_k;y) - \int\limits_0^y dt 
       \left( \frac{\partial}{\partial t} G(z_2,z_1;t) \right) 
       G(z_3,...,z_k;y).
\eq
Repeating this procedure one arrives at the following
integration-by-parts identity:
\bq
\label{ibp}
\lefteqn{
G(z_1,...,z_k;y) + (-1)^k G(z_k,...,z_1;y) } & & \nonumber \\
& = & G(z_1;y) G(z_2,...,z_k;y) - G(z_2,z_1;y) G(z_3,...,z_k;y)
 + ... \nonumber \\
& &
 - (-1)^{k-1} G(z_{k-1},...z_1;y) G(z_k;y),
\eq
which relates the combination $G(z_1,...,z_k;y) + (-1)^k G(z_k,...,z_1;y)$
to $G$-functions of lower depth.
This relation is useful in simplifying expressions.
Eq. (\ref{ibp}) can also be derived in a different way.
In a Hopf algebra we have for any non-trivial element $w$ 
the following relation involving the antipode:
\bq
\label{axiomantipode}
\sum\limits_{(w)} w^{(1)} \cdot {\cal S}( w^{(2)} ) & = & 0.
\eq
Here Sweedler's notation has been used.
Sweedler's notation writes the coproduct of an element $w$ as
\bq
\Delta(w)  & = & \sum\limits_{(w)} w^{(1)} \otimes w^{(2)}.
\eq
Working out the relation (\ref{axiomantipode}) for the shuffle 
algebra of the functions
$G(z_1,...,$ $z_k;y)$, we recover (\ref{ibp}).

We may now proceed and check if (\ref{axiomantipode}) provides
also a non-trivial relation for the quasi-shuffle algebra of $Z$-sums.
This requires first some notation:
A composition of a positive integer $k$ is a sequence $I=(i_1,...,i_l)$ of
positive integers such that $i_1+...i_l = k$.
The set of all composition of $k$ is denoted by ${\cal C}(k)$.
Compositions act on $Z$-sums as 
\bq
\lefteqn{
(i_1,...,i_l) \circ Z(n;m_1,...,m_k;x_1,...,x_k) } & & \nonumber \\
& = & 
 Z\left(n;m_1+...+m_{i_1},m_{i_1+1}+...+m_{i_1+i_2},...,m_{i_1+...+i_{l-1}+1}+...
 \right. \nonumber \\
 & & \left. 
 +m_{i_1+...+i_l}; 
     x_1 ... x_{i_1}, x_{i_1+1} ... x_{i_1+i_2}, ..., x_{i_1+...+i_{l-1}+1} ... x_{i_1+...+i_l}\right),
\eq
e.g. the first $i_1$ letters of the $Z$-sum are combined into one new letter,
the next $i_2$ letters are combined into the second new letter, etc..
With this notation for compositions one obtains the following closed formula for the
antipode in the quasi-shuffle algebra:
\bq
{\cal S} Z(n;m_1,...,m_k;x_1,...,x_k) & = & (-1)^k \sum\limits_{I \in {\cal C}(k)}
 I \circ Z(n;m_k,...,m_1;x_k,...,x_1)
 \nonumber \\
\eq
From (\ref{axiomantipode}) we then obtain
\bq
\lefteqn{
\hspace*{-1cm}
Z(n;m_1,...,m_k;x_1,...,x_k) + (-1)^k Z(n;m_k,...,m_1;x_k,...,x_1) } & & \nonumber \\
 & = & 
 - \sum\limits_{adm. \;cuts} P^C( Z(n;m_1,...,m_k;x_1,...,x_k))
 \nonumber \\
 & &
 \cdot
{\cal S} \left( R^C( Z(n;m_1,...,m_k;x_1,...,x_k)) \right)
\nonumber \\
& &
 - (-1)^k \sum\limits_{I \in {\cal C}(k)\backslash (1,1,...,1) } I \circ Z(n;m_k,...,m_1;x_k,...,x_1).
\eq
Again, the combination 
$Z(n;m_1,...,m_k;x_1,...,x_k) + (-1)^k Z(n;m_k,...,m_1;x_k,$
$...,x_1)$
reduces to $Z$-sums of lower depth, similar to (\ref{ibp}).
We therefore obtained an ``integration-by-parts'' identity for objects, which don't have
an integral representation. 
We first observed, that for the $G$-functions, which have an integral representation,
the integration-by-parts identites are equal to the identities obtained from the antipode.
After this abstraction towards an algebraic formulation, one can translate these relations to cases, which
only have the appropriate algebra structure, but not necessarily a concrete
integral representation.
As an example we have
\bq
\lefteqn{
Z(n;m_1,m_2,m_3;x_1,x_2,x_3) - Z(n;m_3,m_2,m_1;x_3,x_2,x_1) 
 = }
 \nonumber \\
 &  & 
Z(n;m_1;x_1) Z(n;m_2,m_3;x_2,x_3)
- Z(n;m_2,m_1;x_2,x_1) Z(n;m_3;x_3)
 \nonumber \\
 & &
- Z(n;m_1+m_2;x_1 x_2) Z(n;m_3;x_3)
+ Z(n;m_2+m_3,m_1;x_2x_3,x_1)
 \nonumber \\
 & &
+ Z(n;m_3,m_1+m_2;x_3,x_1x_2)
+ Z(n;m_1+m_2+m_3;x_1x_2x_3),
\eq
which expresses the combination of the two $Z$-sums of depth $3$ as $Z$-sums
of lower depth.
The analog example for the shuffle algebra of the $G$-function reads:
\bq
G(z_1,z_2,z_3;y) - G(z_3,z_2,z_1;y) 
 &= &
G(z_1;y) G(z_2,z_3;y)
- G(z_2,z_1;y) G(z_3;y).
\nonumber \\
\eq
Multiple polylogarithms obey both the quasi-shuffle algebra and the shuffle algebra.
Therefore we have for multiple polylogarithms two relations, which are in general
independent.

\section{Summary}
\label{sect:summary}

In this article I discussed the mathematics underlying the calculation 
of Feynman loop integrals. The algorithms are based on $Z$-sums, which form
a Hopf algebra with a quasi-shuffle product.
This algebra has as additional structure a conjugation and a convolution
product.
In the final results multiple polylogarithms appear. Multiple polylogarithms
obey apart from the quasi-shuffle algebra a second Hopf algebra. This 
additional Hopf algebra has a shuffle product.

\begin{appendix}

\section{Notations and conventions}
\label{sect:conventions}

There are several notations for the multiple polylogarithms.
I briefly summmarize them here.
In this article multiple polylogarithms are defined via the sum
representation
\bq
\mbox{Li}_{m_k,...,m_1}(x_k,...,x_1) & = &
Z(\infty;m_1,...,m_k;x_1,...,x_k).
\eq
The reversed order of the arguments and indices 
for $\mbox{Li}_{m_k,...,m_1}(x_k,...,x_1)$ 
follows the notation of Goncharov \cite{Goncharov}.
Gehrmann and Remiddi 
\cite{Gehrmann:2001pz,Gehrmann:2001jv,Gehrmann:2002zr}
use the notation $G(z_1,...,z_k;y)$ and $G_{m_1,...,m_k}(z_1',...,z_k';y)$.
The relation with the notation above is
\bq
\mbox{Li}_{m_k,...,m_1}(x_k,...,x_1)
& = & (-1)^k 
 G_{m_1,...,m_k}\left( b_1, b_2, ..., b_k ;1 \right),
\eq
where
\bq
b_j & = & \frac{1}{x_1 x_2 ... x_j}.
\eq
Borwein, Bradley, Broadhurst and Lisonek \cite{Borwein}
denote multiple polylogarithms as
\bq
\lambda\left(
\begin{array}{c}
m_1,...,m_k \\
b_1,...,b_k \\
\end{array}
\right)
 & = & 
\mbox{Li}_{m_k,...,m_1}(x_k,...,x_1) 
\eq
In the french literature \cite{Minh:2000,Cartier:2001}
harmonic polylogarithms are often denoted as
\bq
\label{french}
\mbox{Li}_{m_1,...,m_k}(x) & = & H_{m_1,...,m_k}(x)  
\eq
and referred to as ``multiple polylogarithms of a single variable''.
Note the order of the indices 
for $\mbox{Li}_{m_1,...,m_k}(x)$ in (\ref{french}).

\end{appendix}


\end{document}